\documentclass{article}

\usepackage{url}
\usepackage{natbib}
\usepackage{graphicx}
\usepackage{amsmath}

\usepackage{listings}
\providecommand{\fftt}{\textit{freefield1010}}

\begin{document}

\title{An open dataset for research on audio \\ field recording archives: \fftt}
\author{Dan Stowell and Mark D. Plumbley\\%
Centre for Digital Music, Queen Mary University of London\\
\texttt{dan.stowell@eecs.qmul.ac.uk}}

\maketitle

\begin{abstract}
We introduce a free and open dataset of 7690 audio clips
sampled from the \textit{field-recording} tag in the Freesound audio archive.
The dataset is designed for use in research
related to data mining in audio archives of field recordings / soundscapes.
Audio is standardised, and audio and metadata are Creative Commons licensed.
We describe the data preparation process,
characterise the dataset descriptively,
and illustrate its use through an auto-tagging experiment.
\end{abstract}

\section{Introduction}
\label{sec:intro}

Digital sound archives hold vast resources of material, including speech, music, and naturalistic and ethnographic field recordings \citep{Ranft:2004}.
However, there are still many challenges in organising and searching in these archives.
In recent decades, research fields such as automatic speech recognition (ASR) and music information retrieval (MIR)
have developed automatic methods for labelling and transcribing specific types of sound.
Yet even if we put speech and music to one side, we still have a large and valuable range of recorded sound,
and there has been relatively little work in organising and searching this non-speech-non-music audio.
Such research may come under the umbrella of ``computational auditory scene analysis'' (CASA)
\citep{Wang:2006}.
Tasks have begun to be addressed in recent years such as automatically labelling the type of audio scene,
or automatically detecting and labelling the events within the audio scene
\citep[ and citations therein]{Giannoulis:2013}.

Important for research development is the existence of standard datasets that can be independently reused by researchers.
This has motivated many community efforts in ASR and in MIR.
In our own recent work we created a set of ``audio scene'' recordings as part of the IEEE AASP ``D-CASE'' challenge,
consisting of 30-second binaural recordings made by three recordists in locations around London
\citep{Giannoulis:2013}.
Such focussed datasets are valuable for developing algorithms for the specific tasks considered,
but they are quite different from most audio archives, being tightly calibrated in their production and of moderate size.

In order for datasets to be relevant to applications in sound archives,
they need to be large enough that they
(a) reflect the diversity of content in audio archives 
and
(b) give some indication of scalability issues for analysis and for visualisation/navigation.
However, large datasets are expensive and time-consuming to record from scratch,
while most existing archives cannot be freely redistributed due to the associated copyright and licensing terms.
It is illustrative that the creators of a recent large pop-music dataset (the \textit{Million Song Dataset}) worked around this issue by distributing not the audio,
but some pre-computed features derived from the audio
\citep{Bertin-Mahieux:2011}.
This enabled them to distribute a large open dataset of relevance to pop music archives,
but as a consequence it restricted the types of analysis possible:
researchers are constrained to using the specific pre-computed features they provided.

For our purposes, a notable initiative is the Freesound archive,%
\footnote{\url{http://freesound.org/}}
which hosts extensive holdings of crowdsourced audio recordings, reusable under Creative Commons and public domain licences.
Established in 2005, it holds more than 160,000 sounds from thousands of users around the world.
It contains a wide range of sound types, including field recordings, recordings from contact mics and hydrophones, and synthetic sounds.

Freesound is already usable for various research purposes, having open licensing conditions
and an easy-to-use application programming interface (API).
However, it is a large crowdsourced and continually-updated collection,
not a fixed dataset.
Files are continually added and removed, and metadata changed;
it is too large to easily redistribute among researchers;
various licences are used, not all mutually compatible;
and
files are in various formats (e.g. WAV/AIFF/ MP3/Ogg, number of channels, sample-rate, duration)
which can be inconvenient for those developing algorithms.
Also, a crowdsourced archive such as Freesound is typically more heterogeneous than traditional archives such as the British Library Sound Archive,
whose curation involves manual attention to file formats and metadata
\citep{Ranft:2004}.

We therefore chose to compile a free and open dataset of a wide range of sounds,
standardised and curated from a fraction of the extensive holdings of the Freesound archive.
This dataset is intended to be of use to researchers developing methods for working with field recordings in audio archives.
In the following we describe how we designed and prepared the data.
We then describe how anyone can access the data and work with it,
and illustrate with an automatic recognition experiment to infer the presence of tags.

\section{Dataset}
\label{sec:data}


In order to maximise the potential usefulness of the dataset, before preparation we considered the following design criteria:

\textbf{Content:} Firstly we aimed to reflect the content that a general audio archive might collect.
Given the range of archive policies, from crowdsourced to strictly curated,
we opted for a middle way, by using Freesound contributions but only those under the \textit{field-recording} tag.
Initial inspection of the various tags in Freesound determined that this tag was mostly free of interpretation issues,
unlike for example \textit{ambience} which Freesound users often use for ambient field recordings but also for synthesised atmospheric soundtrack sounds.
On the other hand, we decided not to manually curate the collection to a specific definition of ``field recording'',
in part because such curation is difficult to apply to material from unknown third parties,
and also because audio archive collections are rarely so narrowly construed.

\textbf{Licensing:} Freesound contains material under various licences.
Most common are the Creative Commons CC-BY licence and the ``CC0'' public domain dedication,
though there is a small proportion of older material under other licences such as Creative Commons ``sampling'' licences
which are not compatible with CC-BY.
We wished to be able to apply a single open licence to the overall dataset,
so we restricted ourselves to CC-BY and CC0 material,
which means the overall dataset can be published under CC-BY.
We also needed to respect the attribution requirements in the CC-BY source material,
which we implemented by ensuring we stored the author metadata with each file as well as a URL to link back to the original source.

\textbf{Size and duration:} We aimed to produce a dataset of manageable storage size, for ease of redistribution, yet suitably diverse.
We also wished to produce sound excerpts of a standard and relatively short duration,
so that they could be used in listening tests without risking listener fatigue,
and so that automatic tests could run efficiently.
These motivations led us to settle on a fixed ten-second duration for each excerpt.

\textbf{File formats:} We also aimed to use a standardised file format.
Freesound allows users to upload sounds of any sample rate, any number of channels,
and in various file formats (uncompressed and lossy-compressed).
We chose not to use audio coming from lossy compressed file formats,
in case of artifacts introduced by the codecs.
We also chose to convert all downloaded sounds to standard CD-quality mono WAV files.
We considered standardising on 24-bit and/or 96 kHz as recommended in 
the IASA ``TC04'' archiving standard \citep{IASA-TC04}.
However, in our experience 16-bit PCM has been more widely compatible than 24-bit PCM:
the latter is not well handled by some older versions of Matlab and some Python audio libraries,
although this situation is improving.
Also the majority of the original downloaded audio was in 44.1 kHz.
We therefore settled on 16-bit 44.1 kHz.
In the crowdsourced Freesound archive, amplitude levels are uncontrolled,
which may be problematic for listening tests.
We therefore chose to amplitude-normalise each excerpt in our dataset.

\textbf{Dataset partitioning:} In data mining and machine-learning experiments, it is useful to have a dataset partitioned into
separate subsets---e.g. one for training and another for testing
\citep[Chapter 5]{Witten:2005}.
To facilitate this we chose to partition our data into ten equally-sized subsets.
Partitioning can be purely random or can be \textit{stratified}:
for example, if the data was intended for an experiment detecting the presence
of geotags, then the partitions could be arranged such that each partition had an equal mix of geotagged and non-geotagged data \citep[Chapter 5]{Witten:2005}.
However, our dataset is intended for various purposes
and not for a single specific experiment,
so we opted for the simple random partitioning.
Researchers can choose to use these subsets for comparability with others,
or to perform their own partitioning.


\subsection{Preparation}

We first obtained specific permission from Freesound to perform our relatively large-scale data download.
We created a Python script based on the official \texttt{freesound-python} code, which we used to download the files from Freesound.%
\footnote{\url{https://github.com/danstowell/freesound-python/tree/tagsearch}}
The script was run to download all files matching all of the following criteria:

\begin{itemize}
	\item	Tagged \textit{field-recording} (which contained 17807 sounds in total)
	\item	Length 10 seconds or greater
	\item	Audio file format WAV
	\item	Published under either the CC-BY licence or CC0
	\item	Audio with 1 or 2 channels
	\item	Audio sample format one of: pcm16, pcm24, pcm32
\end{itemize}
Each file was saved along with its metadata in JSON format.%
\footnote{\url{http://tools.ietf.org/html/rfc4627}}
A small number of files (44) failed to download completely; these were detected by using the \texttt{sox} command-line audio tool (v 14.3.2) to attempt WAV decoding, and deleting the files which reported end-of-file errors or similar.
The script was run in July 2013, taking about a week to download 328 GB of material.

We then prepared a 10-second standardised excerpt from each audio file, taken from the middle of the audio recording, and used \texttt{sox} to convert it to a standardised file format:
WAV, single-channel, sample rate 44.1 kHz, 16-bit PCM, amplitude normalised to $-2$ dB (empirically selected as the maximum gain before clipping).

We inspected the excerpts for any further issues, by listening to all 10-second extracts.
We found a few (seven) which were pure silence with DC offset. While this may sometimes be valid audio when considered in context, it led to normalisation issues, and could be problematic for some applications; we decided to remove these files.

The above procedure resulted in 7690 sound excerpts with accompanying metadata.
We placed the exerpts into 10 separate partitions, where the allocation was by pseudorandom shuffle initialised with a fixed seed value for repeatability (given in the Python script referenced above).
Each of the 10 partitions has about 128 minutes of audio; the dataset totals over 21 hours of audio.

\begin{figure*}[t]
	\centering
	\includegraphics [width=0.7\textwidth,clip,trim=11mm 1mm 20mm 12mm]  {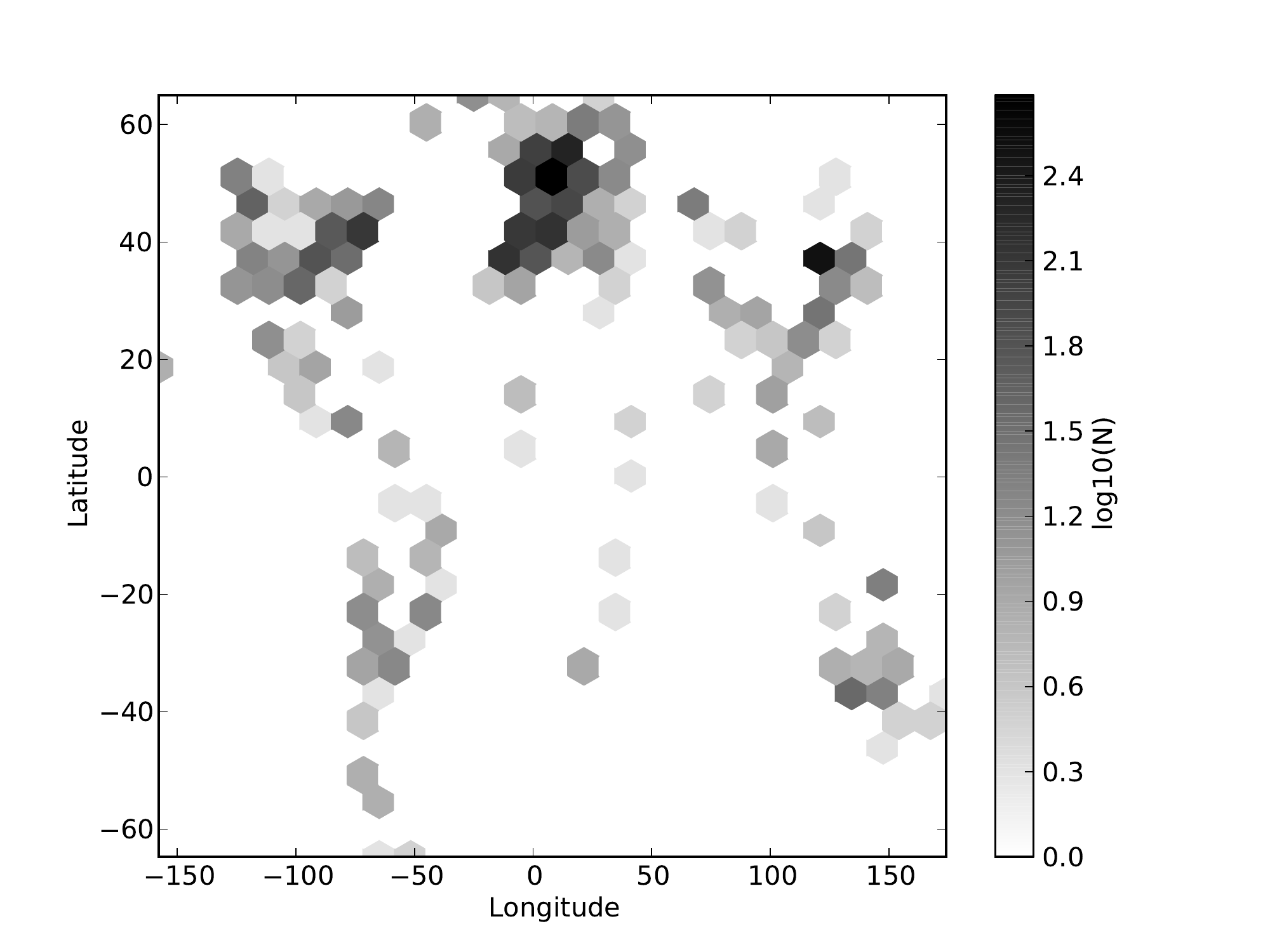}
	\caption{Density plot of all geolocation tags in the dataset. Source code for this plot is in Appendix A.}
\label{fig:plotgeo}
\end{figure*}

Each file is associated with various metadata (author, date, licence etc.) including a median of 7 tags per file (range 1--68).
Around 40\% of the files come with geolocation metadata.
Figure \ref{fig:plotgeo} shows a density plot for all geolocations in the dataset.
It indicates a broad geographic spread, although with much the strongest density in Europe
(perhaps understandable given that Freesound is a European project),
and a relative lack of sounds tagged from Africa and Russia.

\subsection{Availability}
The full dataset is available online, hosted on the Internet Archive,%
\footnote{\url{http://archive.org/details/freefield1010}}
and our institutional repository%
\footnote{\url{http://c4dm.eecs.qmul.ac.uk/rdr/handle/123456789/35}}%
, 
under an overall CC-BY licence.
For portability, each of the ten subsets of the data is presented as a single zip file
which can fit on a data CD (around 570 MB per subset).
We list MD5 checksums of these zip files so that their integrity can be verified.

The metadata stored alongside the audio is easy to work with.
Appendix A gives a simple Python code example for working with the JSON metadata,
in this case the code used to generate Figure \ref{fig:plotgeo}.

\section{A classification experiment}
\label{sec:expt}

To demonstrate the dataset in use,
we conducted an experiment using a binary classification paradigm
using audio content analysis to infer the presence/absence of particular tags (sometimes called \textit{auto-tagging} \citep{Ellis:2011}).
For this we used the simple baseline classifier \textit{smacpy} presented in \citet{Giannoulis:2013},
which provides a standard implementation of a common approach based on Gaussian mixture modelling of Mel frequency cepstral coefficients (MFCCs).

Freesound uses a free-tagging system, with users able to associate an arbitrary number of tags with an audio file,
and also able to create new tags.
We thus manually chose a small selection of tags which appear relatively often---%
\textit{birdsong}, \textit{city}, \textit{people}, \textit{nature}, \textit{train}, \textit{voice}, \textit{water}%
---%
to use for this experiment, creating a binary classification task for each one.
We note that we might expect the tags to have a varying directness of connection with the audio content:
we would expect recordings tagged \textit{voice} to feature human voice sounds relatively prominently,
whereas recordings tagged \textit{city} might contain many sounds in ensemble,
with perhaps no sonic component always present.

We further added two \textit{pseudo-tags} to the study, based on other metadata attributes:
an indicator of whether or not the item comes with geolocation data (\textit{\_\_geotagged}),
and an indicator of whether the item is CC-BY licensed (\textit{\_\_ccby}), as opposed to the only other licence present, public domain.
These pseudo-tags in principle have no direct connection to the audio content,
although there may be correlations due to circumstantial effects
(for example, geolocation might more often be stored when recording outdoor scenes).
We therefore expected only mild if any ability to predict these metadata attributes from audio.

\begin{table}[t]
\caption{Prevalence of selected tags in \fftt.}
\label{tbl:prevalence}
	\centering
\begin{tabular}{l r r}
Tag     	&	Num tagged	&	Proportion (\%)	\\
\hline
birdsong	&	198	&	 2.6	\\
city    	&	562	&	 7.3	\\
nature  	&	905	&	11.8	\\
people   	&	321	&	 4.1	\\
train   	&	411	&	 5.3	\\
voice   	&	556	&	 7.2	\\
water   	&	707	&	 9.2	\\
\_\_geotagged	&	3058	&	39.8	\\
\_\_ccby   	&	6111	&	79.5	\\
\end{tabular}
\end{table}

Table \ref{tbl:prevalence} provides a summary of the prevalence of the selected tags in the \fftt dataset.
It is important to note that, for the ``true'' tags especially, each tag is present in only a minority of the items;
the ratio of positive to negative instances is highly skewed.
This has consequences for how we evaluate automatic classification:
rather than using raw accuracy,
which fails to account for this skew,
we use the area under the curve (AUC) statistic derived from a receiver operating characteristic (ROC) curve
\citep{Provost:1998,Fawcett:2006}.

Our experiment proceeded as follows:
for each of the selected tags and pseudo-tags,
we performed a ten-fold cross-validation experiment using the folds defined by the ten subsets of \fftt.
This means that for each fold,
we used nine of the ten subsets as training data for the classifier,
where the presence/absence of the tag was the binary attribute to be learnt,
and then tested the classifier using the audio from the one remaining subset.
For each such run, we calculated the numbers of correct and incorrect decisions,
and used this to calculate the AUC statistic
(as in \citet{Fawcett:2006}).
Source code for this experiment is available online.%
\footnote{\url{https://github.com/danstowell/smacpy/tree/freefield1010}}

\begin{figure}[t]
	\centering
	\includegraphics [width=0.5\textwidth,clip,trim=11mm 1mm 20mm 12mm]  {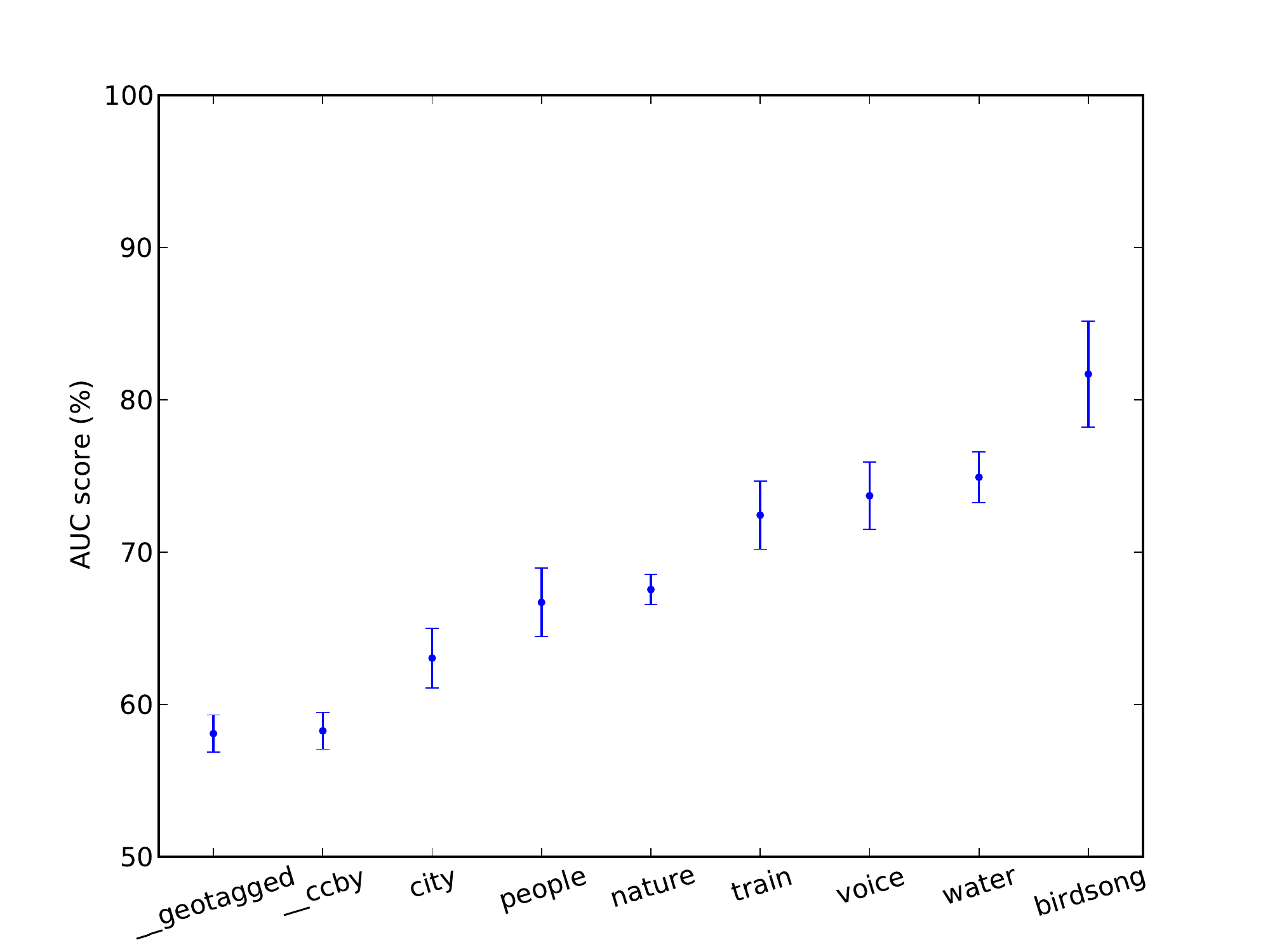}
	\caption{Results for automatic inference of tag presence/ absence. %
	Plot shows the mean and the 95\% confidence interval of the AUC score for each tag (and pseudo-tag) studied, across ten-fold crossvalidation.
	}
\label{fig:plotff1010smacpyclassif}
\end{figure}

Results in Figure \ref{fig:plotff1010smacpyclassif}
show that the tags can be automatically inferred from audio with varying degrees of reliability.
The best result is for \textit{birdsong} at 82\% AUC,
while the weakest result for a true tag is \textit{city} at 63\%.
Note that the standard interpretation of an AUC value is
that it tells us the probability that the algorithm will rank a random positive instance
higher than a random negative instance \citep{Fawcett:2006};
chance performance is always 50\% for the AUC statistic.
Results for the two pseudo-tags attain around 58\% AUC:
above chance,
but still very weak.
It indicates that there is some mild difference in audio content between the positive and negative instances,
but much less than for the true tags.

The classification performance for true tags appears to show some connection with the ``directness'' issue raised above:
the tags which yield weakest recognition performance (\textit{city}, \textit{people}, \textit{nature})
can be said to have an indirect connection with the audio content.
However these results are illustrative only, using a baseline classifier rather than a leading-edge algorithm.
The strongest performance (82\%) is still much lower than is desirable for a binary classifier deployed in a live system.

The 95\% confidence intervals (error bars) in Figure \ref{fig:plotff1010smacpyclassif} are relatively small and well-separated.
This illustrates that the dataset is of sufficient size to make inferences about the relative predictability of these tags from the audio content,
and also that there is relative consistency among the ten folds of the data.

\section{Conclusions}
\label{sec:conc}

In this paper we have described the preparation of a free and open audio dataset,
designed primarily for use in research on data mining of audio archives.
The dataset is derived from a subset of the Freesound archive,
but fixed and standardised so as to facilitate reproducible research.
In an auto-tagging experiment we demonstrated that the dataset can be used to probe issues
such as the differential predictability of tags from audio.
We hope that the dataset will prove useful to others.

\section*{Acknowledgments}

We wish to thank the Freesound developers and maintainers at the Music Technology Group in Universitat Pompeu Fabra,
for running the excellent Freesound archive and for allowing us API access to the data.

DS \& MP are supported by an EPSRC Leadership Fellowship EP/G007144/1.

\bibliographystyle{plainnat}
\bibliography{../refs}

\clearpage
\onecolumn
\section*{Appendix A: Code example}
\label{sec:codeexample}
The following Python code was used to generate Figure \ref{fig:plotgeo} (using Python 2.7).

\begin{lstlisting}[language=Python]
import glob, json
import matplotlib.pyplot as plt
import matplotlib.cm as cm

ffpath = './freefield1010'
outfile = 'plots/plotgeo.pdf'

lats = [], lons = []

for onepath in glob.iglob('%s/*/*.json' % ffpath):
	jsonfile = open(onepath, 'r')
	jsondata = json.load(jsonfile)
	jsonfile.close()
	geo = jsondata.get(u'geotag')
	if geo != None:
		lats.append(float(geo[u'lat']))
		lons.append(float(geo[u'lon']))

# plot
plt.figure()
plt.hexbin(lons, lats, gridsize=25, bins='log', cmap=cm.binary)
plt.xlabel('Longitude')
plt.ylabel('Latitude')
cb = plt.colorbar()
cb.set_label('log10(N)')

plt.savefig(outfile, papertype='A4', format='pdf')

\end{lstlisting}
\end{document}